\documentclass[sigconf]{acmart}
\usepackage{multirow}
\usepackage{booktabs,tabularx,subcaption}
\usepackage{algorithm}
\usepackage{algpseudocode}

\algrenewcommand\algorithmicrequire{\textbf{Input:}}
\algrenewcommand\algorithmicensure{\textbf{Output:}}

\AtBeginDocument{%
  }

\renewcommand\footnotetextcopyrightpermission[1]{}
\acmSubmissionID{1359}

\begin{document}

\title{IncreRTL: Traceability-Guided Incremental RTL Generation under Requirement Evolution}

\author{ Luanrong Chen}
\affiliation{%
  \institution{ National University of Defense Technology}
  \city{ Changsha}
  \country{ China}
}
\email{ chenluanrong@nudt.edu.cn}

\author{ Renzhi Chen}
\affiliation{%
  \institution{Defense Innovation Institute, Academy of Military Sciences}
  \city{Beijing}
  \country{China}
}
\affiliation{%
  \institution{Qiyuan Lab}
  \city{Beijing}
  \country{China}
}

\author{ Xinyu Li}
\affiliation{%
  \institution{National University of Defense Technology}
  \city{Changsha}
  \country{China}
}

\author{ Shanshan Li}
\affiliation{%
  \institution{National University of Defense Technology}
  \city{Changsha}
  \country{China}
}

\author{ Rui Gong}
\affiliation{%
  \institution{National University of Defense Technology}
  \city{Changsha}
  \country{China}
}

\author{ Lei Wang}
\authornotemark[1]
\affiliation{%
  \institution{Defense Innovation Institute, Academy of Military Sciences}
  \city{Beijing}
  \country{China}
}
\affiliation{%
  \institution{Qiyuan Lab}
  \city{Beijing}
  \country{China}
}

\renewcommand{\shortauthors}{}

\begin{abstract}
  Large language models (LLMs) have shown promise in generating RTL code from natural-language descriptions, but existing methods remain static and struggle to adapt to evolving design requirements, potentially causing structural drift and costly full regeneration. We propose IncreRTL, a LLM-driven framework for incremental RTL generation under requirement evolution. By constructing requirement–code traceability links to locate and regenerate affected code segments, IncreRTL achieves accurate and consistent updates. Evaluated on our newly constructed EvoRTL-Bench, IncreRTL demonstrates notable improvements in regeneration consistency and efficiency, advancing LLM-based RTL generation toward practical engineering deployment.
\end{abstract}





\maketitle


\section{Introduction}
Register-Transfer Level (RTL) design is a critical stage in chip development~\cite{yang2023back, yang2022unicorn, ma2024darwin3}, serving as the bridge between high-level specifications and low-level circuit implementations. The rapid advancement of large language models (LLMs) has introduced new opportunities for automating RTL hardware design. By leveraging their strong capabilities in semantic understanding and code synthesis, LLMs can generate Verilog code from natural-language specifications, thereby significantly enhancing design productivity.

\begin{figure}[t]
  \centering
  \includegraphics[width=1\linewidth]{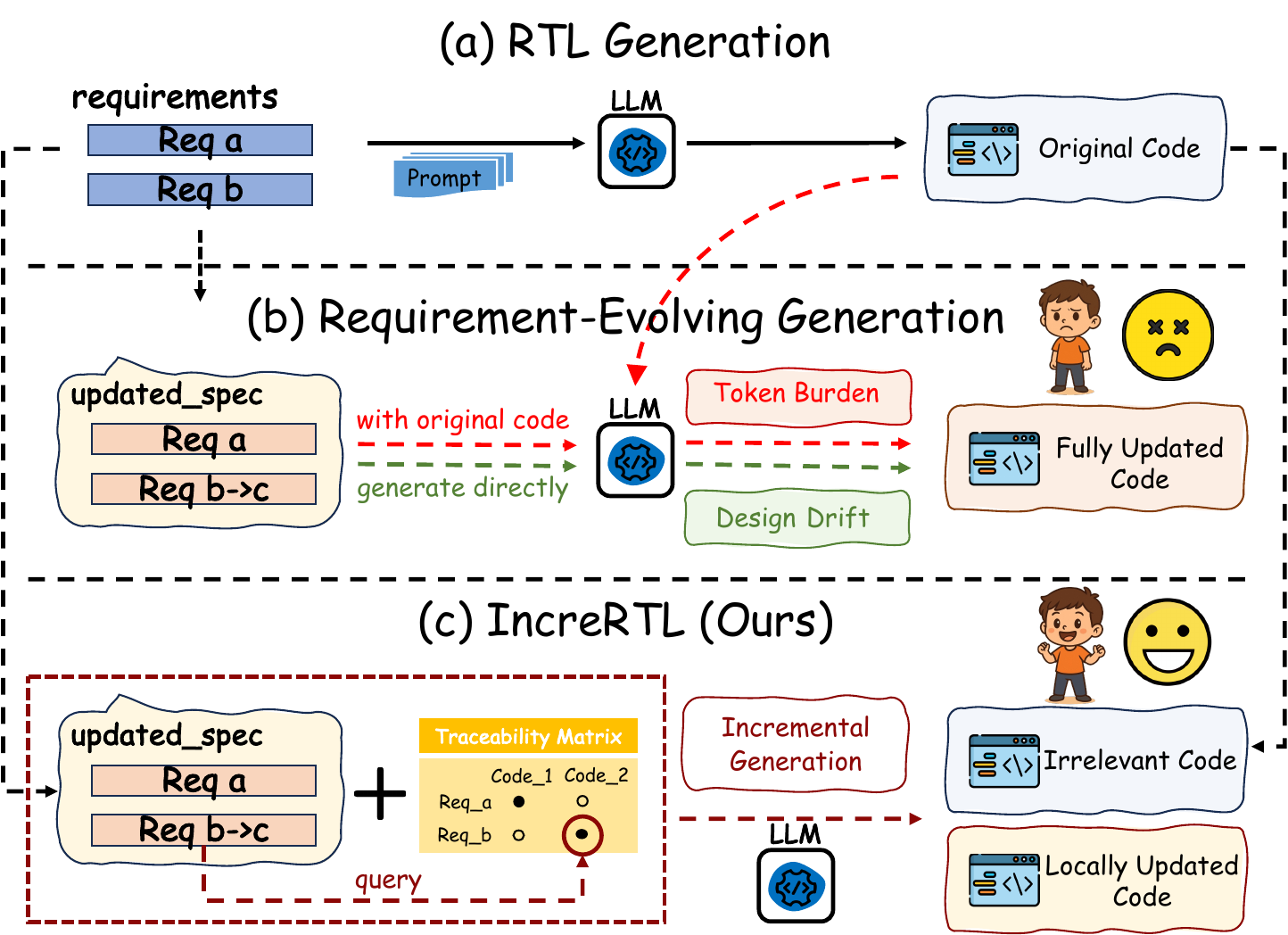}
  \caption{Illustration of requirement-driven RTL updates. IncreRTL performs traceability-guided local regeneration, mitigating token and drift issues.}
  \Description{Illustration showing requirement-driven modification in LLM-based RTL generation, including standard and evolved workflows.}
  \label{fig:change}
\end{figure}

\begin{figure*}[t]
  \centering
  \includegraphics[width=1\textwidth]{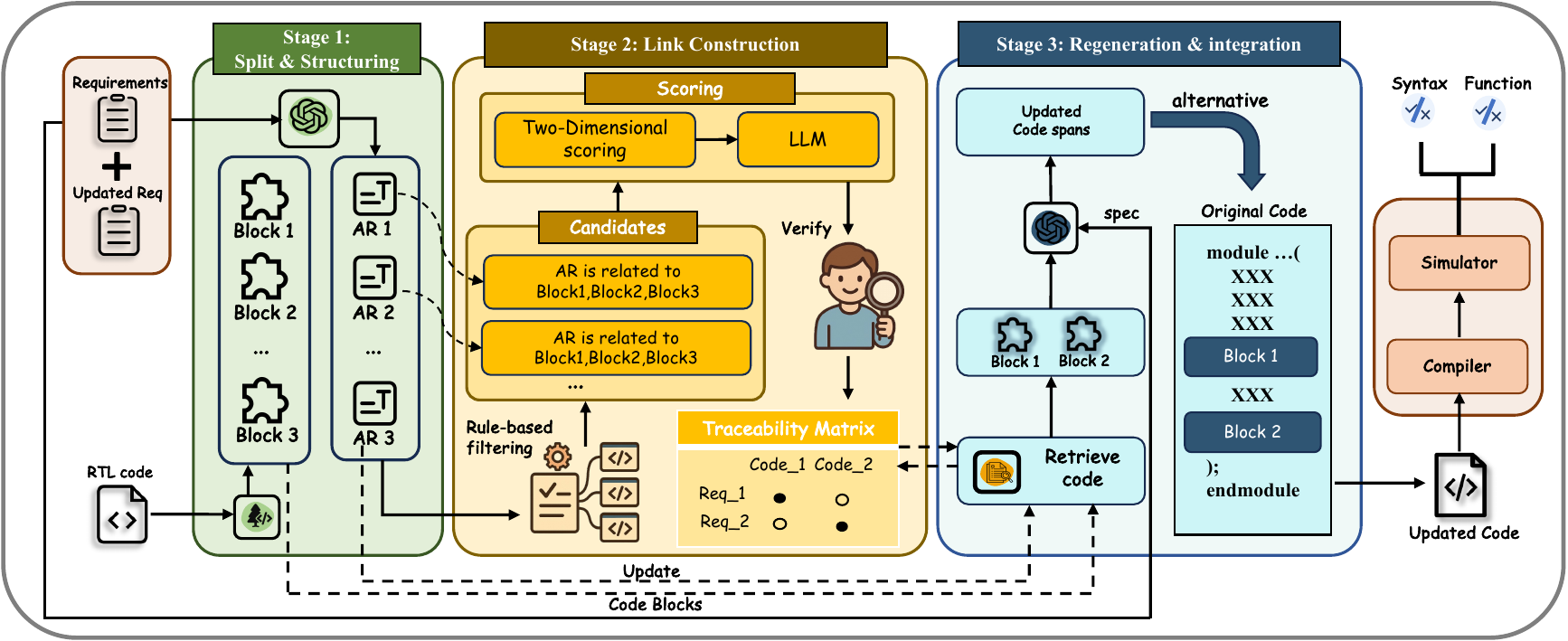}
  \caption{Overview of the proposed IncreRTL framework.}
  \Description{Overview diagram showing the architecture and workflow of IncreRTL.}
  \label{fig:overview}
\end{figure*}

However, existing LLM-driven RTL code generation approaches remain limited in their ability to accommodate continuously evolving design requirements. In practical hardware development, such evolution is inevitable, typically triggered by specification updates, functional extensions, or performance adjustments. Prior studies~\cite{thakur2024verigen, yao2024rtlrewriter, chang2024data, chang2023chipgpt} have predominantly evaluated generation quality under static specifications, emphasizing one-shot code generation as indicated in (a) of Figure~\ref{fig:change}. Even approaches that leverage Multi-Agent frameworks~\cite{zhao2025mage, gao2024autovcoder} for iterative refinement still assume unchanged inputs. Our observations indicate that even minor requirement modifications can cause regenerated code to diverge markedly from the original implementation as shown by the red dashed arrows in (b) of Figure~\ref{fig:change}, often forcing downstream engineering efforts to restart from scratch. Providing the original implementation together with the updated requirement description can partially suppress this design drift as shown by the green dashed arrows in (b) of Figure~\ref{fig:change}. However, this approach leads to a significant increase in prompt length, which in turn raises the inference overhead. Moreover, prior work~\cite{leng2024long} suggests that excessively long contextual prompts dilute salient information. As the complexity of the design increases, this effect becomes more significant, ultimately leading to a degradation in generation quality.

To address these challenges, it is essential to establish a mechanism that preserves consistency and interpretability across iterative updates. In engineering practice, requirement traceability~\cite{tufail2017systematic} has long been recognized as an effective means of managing the evolution of design specifications. By systematically constructing traceability links between requirements and their corresponding implementations, such traceability enables consistent change management throughout the design lifecycle. Recent advances in LLM-based semantic understanding further create new opportunities for automating the construction of these links. However, this paradigm remains underexplored in hardware design. Because LLM generation inherently depends on its input, requirement descriptions that lack explicit pointers to implementation locations often lead to global drift triggered by local changes. The structured representation of RTL provides a natural basis for anchoring requirements to their implementations, and leveraging LLM semantic capabilities to construct traceability links that introduces the necessary structural guidance into model inputs, thereby enabling localized and consistent updates across design iterations.

Motivated by these observations, we introduce \textbf{IncreRTL}, the first LLM-driven RTL generation framework specifically designed to address evolving design specifications. This framework establishes requirement--code traceability links to precisely locate and incrementally regenerate affected code fragments, ensuring structural consistency while minimizing the overhead of full regeneration. As shown in \textbf{Figure~\ref{fig:change}(c)}, IncreRTL combines the semantic understanding and logical reasoning capabilities of large language models with chain-of-thought (CoT) inference ~\cite{wei2022chain} to create interpretable correspondences and capture the impact of specification changes on the entire design. To rigorously evaluate the framework, we developed \textbf{EvoRTL-Bench}, the first benchmark specifically tailored for assessing LLM-driven RTL regeneration under evolving requirements. Experimental results on EvoRTL-Bench demonstrate that IncreRTL significantly outperforms two baseline methods in consistency, with full-code regeneration requiring \textbf{23.29\%} more tokens than our approach. Moreover, IncreRTL consistently achieves more efficient and stable RTL regeneration across various requirement changes and large language models, showcasing its scalability and generalization capabilities in real-world design scenarios.\\

In summary, our main contributions are as follows:
\begin{itemize}
    \item We propose \textbf{IncreRTL}, the first LLM-driven framework for requirement-aware RTL generation under evolving design specifications.
    \item We introduce a method that leverages LLMs to construct bidirectional semantic alignments between RTL code and requirement descriptions, enabling localization and incremental regeneration.
    \item We introduce \textbf{EvoRTL-Bench}, the first systematic benchmark designed to evaluate the performance of large language models in RTL generation under evolving design requirements.
\end{itemize}

\section{Related Work}
\subsection{Large Language Models for RTL Code Generation}
DAVE~\cite{pearce2020dave} pioneered LLM-based NL-to-HDL translation, while VeriGen~\cite{thakur2024verigen} and RTLCoder~\cite{liu2024rtlcoder} enhance correctness via large-scale Verilog-specific training. VToT~\cite{zhou2025vtot} introduces a Verilog-specific Tree-of-Thoughts prompting scheme to improve LLM-based RTL generation, and ChipNeMo~\cite{liu2023chipnemo} enhanced generalization via EDA-specific training. Despite these advances, most approaches still rely on static one-shot generation and fail to adapt to evolving design requirements. Recent iterative methods, such as RTLFixer~\cite{tsai2024rtlfixer} and VerilogCoder~\cite{ho2025verilogcoder}, leverage compiler diagnostics and simulation feedback to enhance correctness, while ARSP~\cite{yao2025arsp} employs semantics-guided fragmentation with a two-stage LLM pipeline to debug and repair Verilog designs at the fragment level. However, these techniques primarily focus on post-hoc error correction rather than managing requirement evolution. To address this gap, we incorporate requirement-to-implementation traceability into the generation pipeline, facilitating the effective management of requirement changes and ensuring consistent updates to the RTL.


\subsection{Requirement Traceability Links}
Existing studies on requirement traceability have addressed the challenge of establishing semantic links between requirements and implementation artifacts. Early IR-based approaches~\cite{antoniol2002recovering, marcus2003recovering} relied on lexical or statistical similarity to infer trace links. Subsequent studies~\cite{guo2017semantically, dai2023constructing} introduced neural semantic models to capture deeper contextual relationships, improving link accuracy and interpretability. More recently, LLM-driven frameworks such as LiSSA~\cite{fuchss2025lissa} and TVR~\cite{niu2025tvr} leverage retrieval-augmented generation (RAG) together with reasoning-based traceability link recovery to establish interpretable cross-artifact links. 

Despite these advances, existing research has primarily focused on software engineering, where traceability techniques have matured and been effectively integrated into requirement management workflows. In contrast, their adaptation to hardware design automation remains largely unexplored. To address this gap, we investigate how to construct traceability links between requirement and Verilog code in RTL design, enabling consistent alignment between evolving specifications and their hardware implementations.

\begin{figure}[t]
  \centering
  \includegraphics[width=\linewidth]{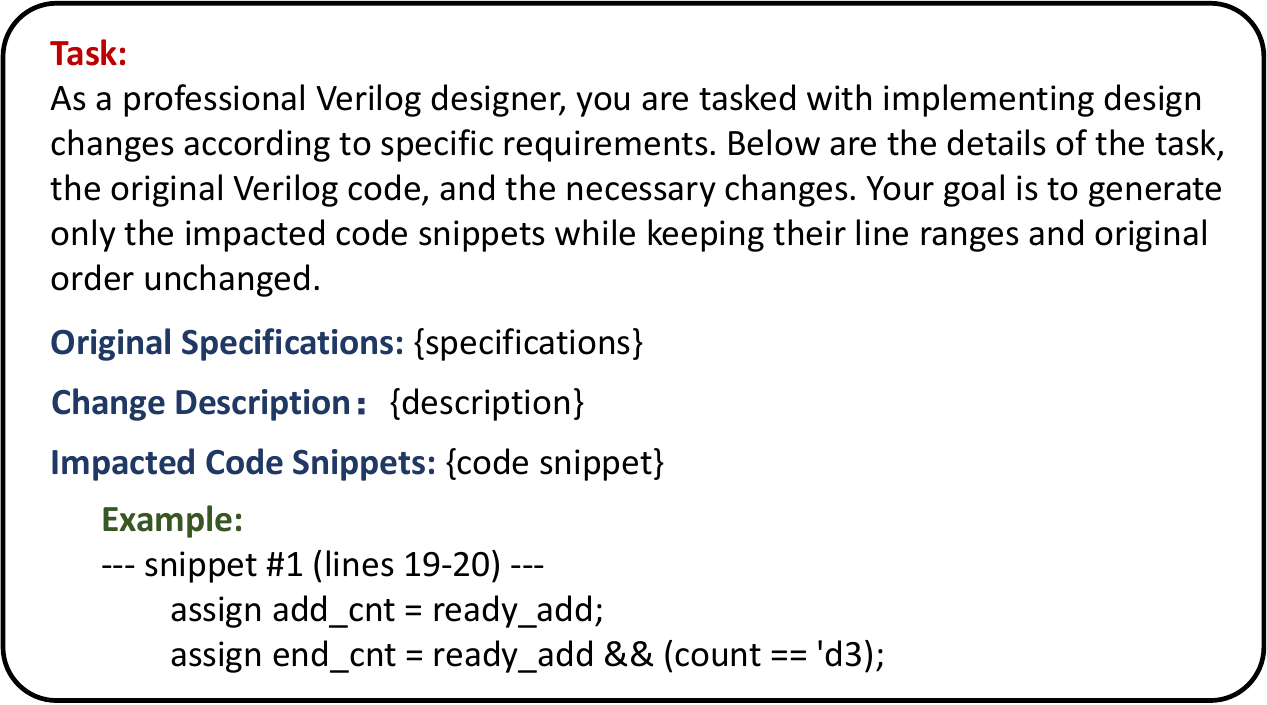}
  \caption{Task template for incremental generation, where the LLMs generate only the impacted code snippets, ensuring unchanged line ranges and interfaces.
}
  \Description{Illustration showing requirement-driven modification in LLM-based RTL generation, including standard and evolved workflows.}
  \label{fig:prompt template}
\end{figure}

\section{Methodology}


We propose IncreRTL, a framework that leverages large language models to support RTL generation under continuously evolving design requirements.
\subsection{Overview}

 The framework takes both the original and updated requirement descriptions, along with the corresponding implementation code, as input to generate RTL designs that align with the latest specifications as shown in Figure \ref{fig:overview}. The process begins with a large language model extracting high-level semantic representations of the requirements, followed by a parser that segments the source code into syntax-preserving blocks. Leveraging these structured representations, candidate links are first identified through rule-based matching based on hierarchical relationships, and then scored across textual and semantic dimensions. The LLMs further refine these links by identifying and filling in any missing connections from a higher level semantic perspective. After validation, the final traceability matrix is obtained. Guided by this matrix, IncreRTL first generates the complete RTL code, then performs localized updates to the affected code fragments, and finally reconstructs and validates the code through compilation and testbench simulation.


\subsection{Traceability Link Construction}
This stage builds semantic alignments between requirements and Verilog code to generate a traceability matrix for subsequent incremental generation and integration.

\subsubsection{Structured Representation Construction}
\paragraph{\textbf{Syntax-Preserving Code Segmentation}}

To construct traceable implementation units, we leverage a Verilog parser to automatically identify the syntactic structure of the code and partition the design into syntax-preserving blocks that reflect hierarchical relationships. 
These blocks correspond to well-defined structural code regions, such as port declarations, register declarations, and combinational logic, ensuring that local updates minimize the disruption of syntactic boundaries.

\begin{algorithm}[t]
\caption{Enhanced Link Scoring and Validation}
\label{alg:enhanced_link_scoring_validation}
\begin{algorithmic}[1]
\Require Requirement set $R$, Code set $C$, Candidate links $L_{\text{cand}}$
\State \textbf{Parameters:}  $\theta_{\text{agg}} = 0.6$
\State Initialize empty score maps $S_{\text{lex}}$, $S_{\text{sem}}$, $S_{\text{agg}}$; $L_{\text{initial}} \gets \emptyset$

\For{each link $(r_i, c_j) \in L_{\text{cand}}$}
    \State $K_r \gets \text{ExtractKeywords}(r_i)$
    \State $K_c \gets \text{ExtractKeywords}(c_j)$
    \State $S_{\text{lex}}(r_i, c_j) \gets \dfrac{|K_r \cap K_c|}{|K_r \cup K_c|}$

    \State $\mathbf{v}_r \gets \text{CodeBERT}(r_i)$
    \State $\mathbf{v}_c \gets \text{CodeBERT}(c_j)$
    \State $S_{\text{sem}}(r_i, c_j) \gets \text{CosineSimilarity}(\mathbf{v}_r, \mathbf{v}_c)$
    \State $S_{\text{agg}}(r_i, c_j) \gets S_{\text{lex}}(r_i, c_j) + S_{\text{sem}}(r_i, c_j)$
    
    \If{$S_{\text{agg}}(r_i, c_j) \geq \theta_{\text{agg}}$}
        \State $L_{\text{initial}} \gets L_{\text{initial}} \cup \{(r_i, c_j)\}$
    \EndIf
\EndFor

\State $L_{\text{missing}} \gets \text{LLM\_Reasoning}(R, C, L_{\text{initial}}, S_{\text{agg}})$
\State $L_{\text{final}} \gets L_{\text{initial}} \cup L_{\text{missing}}$
\State $\mathbf{M} \gets \text{ManualValidation}(L_{\text{final}})$

\Ensure Traceability matrix $\mathbf{M}$
\end{algorithmic}
\end{algorithm}

\paragraph{\textbf{Semantic Requirement Representation Construction}}



We leverage the Chain-of-Thought (CoT) technique~\cite{wei2022chain} to steer large language models (LLMs) through a structured reasoning process by providing explicit guidance within the input prompts. Dedicated prompt templates are used to help the model grasp both the high-level requirement description and the decomposed atomic requirements, where each atomic requirement represents the smallest functional unit derived from the overall specification. Each atomic requirement is subsequently analyzed from the RTL implementation perspective, and the 
resulting structured representation captures five core elements: \textit{interfaces}, \textit{signals}, \textit{trigger conditions}, \textit{behaviors}, and \textit{state transitions}.

\begin{figure}[t]
  \centering
  \includegraphics[width=1\linewidth]{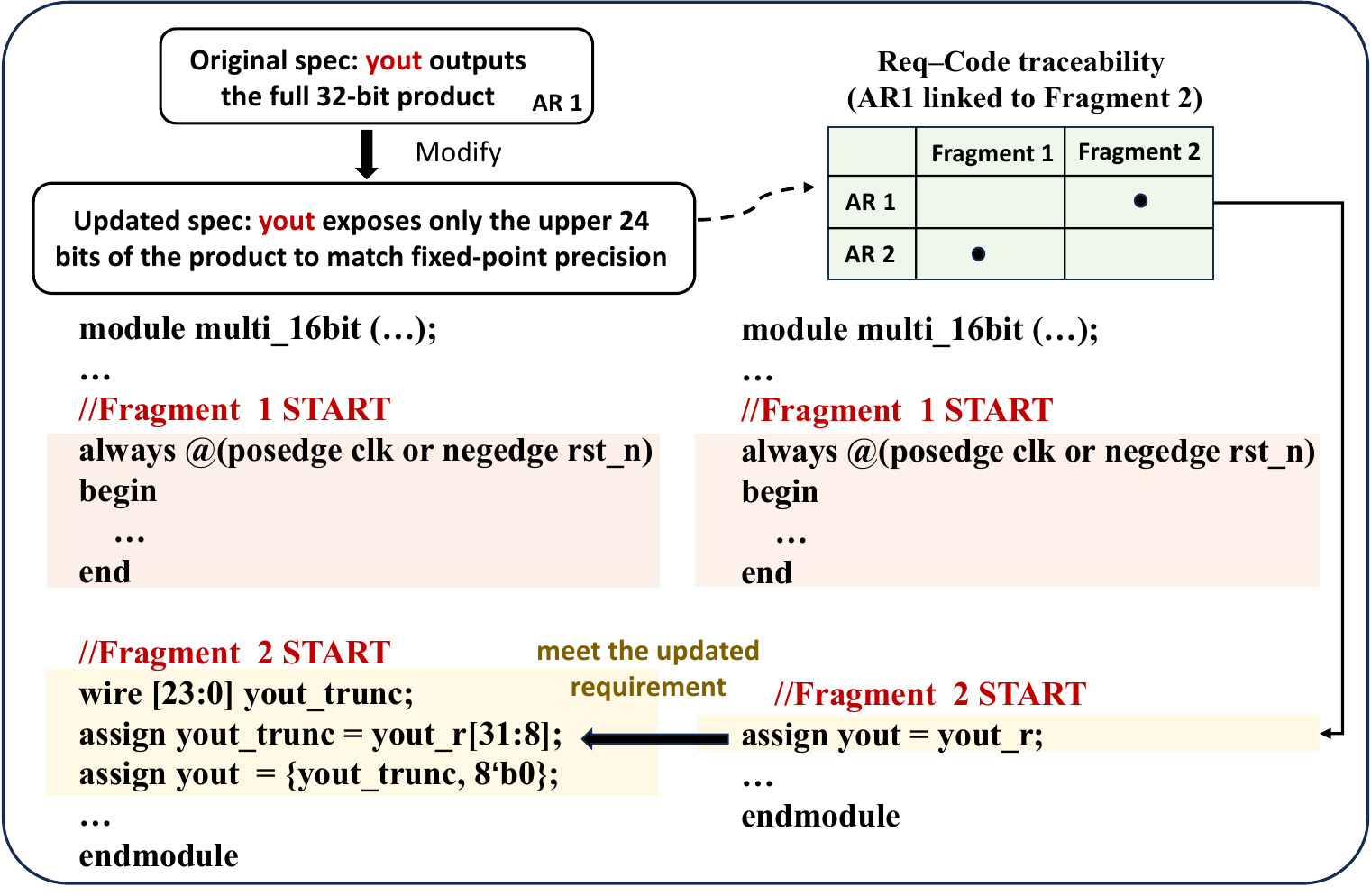}
  \caption{Example of localized regeneration for AR1, where the requirement changes from full 32-bit output to exposing only the upper 24 bits; traceability links confine updates to Fragment 2, and IncreRTL regenerates only its output logic while keeping Fragment 1 unchanged.}
  \Description{Illustration showing requirement-driven modification in LLM-based RTL generation, including standard and evolved workflows.}
  \label{fig:code}
\end{figure}

\subsection{Trace-Guided Incremental Updates}

\subsubsection{Requirement--Code Semantic Alignment}
\paragraph{\textbf{Candidate Links Based on Hierarchical Relationships}}

To generate candidate links, we establish a concise correspondence 
between the structured atomic requirement fields and the hierarchical 
layers of the Verilog implementation. 

Under this alignment,
\textit{interfaces} are associated with the module-header layer, 
\textit{signals} with the declaration layer, \textit{trigger conditions} 
with conditional constructs appearing in continuous, combinational, 
or sequential logic layers, \textit{behaviors} with functional logic within combinational or sequential layers, and \textit{state 
transitions} with the finite-state-machine layer. Based on this correspondence, we obtain candidate links by selecting 
Verilog code blocks from the layers associated with the fields of each atomic requirement.


\paragraph{\textbf{Link Scoring and Validation}}
In this step, we calculate scores for each requirement-to-code candidate link based on two main dimensions: lexical matching and semantic matching.

\begin{enumerate}
    \item \textbf{Lexical Matching} identifies keyword overlaps between the requirement and the Verilog code, ensuring that code elements referenced in the requirement are correctly linked. Compared with the preceding structural filtering stage, it offers a finer-grained assessment of link relevance. Similarity is quantified using the Jaccard coefficient~\cite{niwattanakul2013using}, computed over keyword sets extracted from structured representations of both the requirement and the code.

    \item \textbf{Semantic Matching} captures deeper relationships. It maps the requirement and the code into a shared semantic vector space using a pre-trained embedding model, and cosine similarity is computed to measure their semantic alignment.

\end{enumerate}

As detailed in Algorithm~\ref{alg:enhanced_link_scoring_validation}, the specific procedures for computing the lexical and semantic scores are explicitly defined. After obtaining these two scores, they are aggregated to produce a consolidated score for each candidate link. 

Subsequently, a pre-trained large language model is employed for high-level semantic reasoning. The complete requirement description, along with the candidate links and their associated scores, is input into the model, which evaluates the requirement-code relationships from a broader semantic perspective and identifies additional semantically relevant links that may have been overlooked during initial retrieval. These supplementary links are incorporated into the evolving traceability matrix. Finally, all generated links undergo further manual validation to ensure their accuracy and relevance, ultimately forming the definitive requirement-to-code traceability matrix.

\begin{table*}[t]
  \caption[Unified overview of requirement changes in \emph{EvoRTL-Bench}.]{
    Sample examples of different requirement change types and modules in \emph{EvoRTL-Bench}, showing the magnitude and count for each type.
  }
  \label{tab:bench}
  \centering
  \footnotesize
  \setlength{\tabcolsep}{4pt}
  \renewcommand{\arraystretch}{1.1}

  \begin{tabularx}{0.85\textwidth}{@{}l l X c l@{}}
    \toprule
    \textbf{Magnitude} & \textbf{Category} & \textbf{Change Description} & \textbf{Count} & \textbf{Module} \\
    \midrule

    \multirow{4}{*}{Minor}
      & FunctionalBehaviorChange 
      & Add an enable input and an output for detecting any edge 
      & 34 & Edge Detector \\

      & ControlConfigChange
      & Add an enable signal to pause/resume the counter
      & 28 & Counter \\

      & InterfaceProtocolChange
      & Add a second independent address and data port for dual-read
      & 8 & ROM  \\

      & MicroArchitectureRefactor
      & Replace serial iteration with parallel array multiplication
      & 20  & Multiplier \\

    \midrule

    \multirow{2}{*}{Major}
      & ModuleStructuralRefactor
      & Split arithmetic and logic units into separate submodules
      & 22 & ALU \\

      & SystemInterfaceRefactor
      & Replace external interface with standard AXI bus protocol
      & 8 & Memory Controller \\



    \bottomrule
  \end{tabularx}

\end{table*}

Building upon the established requirement–traceability link matrix, this stage performs requirement-driven incremental generation and localized in-place updates of RTL fragments. To support this process, we introduce a task-oriented prompt template, as shown in Figure~\ref{fig:prompt template}, comprising three core components: the original requirement description, the requirement modification statement, and the set of relevant code snippets retrieved from the traceability links. Each snippet is annotated with its corresponding line number in the source file and organized into an explicitly indexed sequence, enabling the model to accurately follow the prescribed snippet order during generation.

In this task, the large language model acts as an RTL reconstruction assistant, performing localized rewrites only for the affected code regions based on the provided snippets. It is instructed to update the implementation exclusively for these snippets, without altering unaffected portions of the design. To ensure the integrity of the incremental reconstruction, the model preserves the original snippet indices and line numbers, which are crucial for maintaining contextual coherence and avoiding structural drift. After generating the updated snippets, a line-offset–based merging mechanism is employed to reintegrate them into the source code. The system computes offsets between the original and updated line numbers, adjusting the insertion positions of subsequent snippets as necessary. This ensures that all updated snippets are precisely placed without misalignment or unintended overwriting. As shown in Figure \ref{fig:code}, the model regenerates and replaces the original portions within the syntactically preserved fragments.

This approach differs from methods that regenerate entire code by updating only the affected fragments, maintaining high consistency with the original design throughout the iterative process. The localized generation method reduces reliance on large-scale context, minimizes token usage, and improves modification efficiency. After each iteration, the system validates the updated fragments through syntax checking with the Verilog compiler and functional simulation using a Testbench, ensuring both syntactic accuracy and functional reliability.

\section{Experiment}
\subsection{Experimental Setup}

All experiments are conducted using the DeepSeek-V3~\cite{liu2024deepseek} via API access, except for the generalization experiments, which are performed on other LLMs.  Detailed information is provided below.

\subsubsection{Benchmark Construction}



To evaluate our method’s performance in handling requirement changes, we construct a new benchmark, EvoRTL-Bench, derived from the RTLLM~\cite{lu2024rtllm} corpus. EvoRTL-Bench is specifically designed to systematically assess the efficiency, consistency, and overall accuracy of code generation quality under diverse evolving design requirements.

All requirement-change instances are created by first decomposing each module's original requirement into atomic units and then using a large language model to generate corresponding requirement modifications and revised Verilog implementations. Variants that successfully compile but fail the original testbench are deliberately retained, since they capture nontrivial behavioral deviations from the original specification and are therefore more informative for evaluating requirement-aware regeneration.

The resulting benchmark comprises 30 design modules and 120 requirement-change instances, spanning both minor and major modification types. Table~\ref{tab:bench} summarizes six representative categories of requirement changes, including two major and four minor types, together with the number of instances in each category. Each instance includes a natural-language requirement description, a change specification, the original Verilog implementation, and a functional testbench. During evaluation, the model regenerates Verilog code conditioned on the requirement-change pair, optionally augmented with the original code as additional context. The generated outputs are then verified through compilation and simulation to assess both syntactic correctness and functional behavior.

\subsubsection{Evaluation Metrics}
This section introduces the evaluation metrics used to measure the performance of large language model--driven RTL generation under evolving design requirements.

\paragraph{\textbf{Consistency Score}}
We define a Consistency Score (CS), which is computed as the normalized similarity across two dimensions: Interface Consistency (ICS) and Structural Consistency (SCS). 

\begin{enumerate}
    \item \textbf{Interface Consistency (ICS)}: The port definitions in the original and generated code are compared, including names, types, widths, and order. Port name consistency is evaluated using the Levenshtein distance, while other inconsistencies are assigned a value of 0.
    
    \item \textbf{Structural Consistency (SCS)}: The Verilog code is parsed into an Abstract Syntax Tree (AST), and the tree edit distance algorithm is used to measure consistency.
\end{enumerate}

The final score is synthesized as follows:
\begin{equation}
CS =   ICS +  SCS
\end{equation}

\begin{table}[t]
  \caption{Comparison of IncreRTL with Direct and Full baselines on the EvoRTL benchmark.
  The table reports Consistency Score (CS), Relative Token Usage (RTU), and syntactic and functional correctness (pass@k).}
  \label{tab:RQ1}
  \centering
  \footnotesize
  \resizebox{\columnwidth}{!}{%
    \begin{tabular}{lcccccc} 
      \toprule
      Method & Consistency & Relative Token & \multicolumn{2}{c}{Syntax} & \multicolumn{2}{c}{Function} \\
      \cmidrule(lr){4-5}\cmidrule(lr){6-7}
             & Score       & Usage      & pass@5 & pass@10   & pass@5 & pass@10 \\
      \midrule
      Direct          & 0.2684  &   -   & 78.15 & 80.03 & 56.30 & 60.00 \\
      Full            & 0.7337  &  1.8   & 89.17 &  90.01 & 67.67 & 70.33 \\
      Ours & 0.8123  &   1.46   & 89.08 & 89.17 & 61.34  & 64.17 \\
      \bottomrule
    \end{tabular}%
  }
\end{table}

\paragraph{\textbf{Relative Token Usage}}
To quantify the computational overhead introduced by prompt extension, we define the Relative Token Usage (RTU) as the ratio between the total tokens consumed by a given method and those consumed by the baseline direct prompting:

\begin{equation}
  RTU = \frac{T_{\text{prompt}} + T_{\text{response}}}
              {T_{\text{prompt}}^{\text{base}} + T_{\text{response}}^{\text{base}}}.
\end{equation}

\paragraph{\textbf{Correctness Metrics}}

We evaluate both syntactic and functional correctness using the pass@$k$ metric, which measures the probability that at least one of the top-$k$ candidates satisfies compilation and testbench constraints.

\begin{equation}
\text{pass@}k = \mathbb{E}_i \left[ 1 - \frac{\binom{n - c_i}{k}}{\binom{n}{k}} \right],
\end{equation}

\subsubsection{Baselines}

We evaluate \textbf{IncreRTL} by comparing it with two baseline settings, all implemented on the DeepSeek-V3 model:
\begin{enumerate}
    \item \textit{Direct Generation}: the model takes only the change description as input and directly produces the Verilog code.
    \item \textit{Full-Code Generation}: the model receives both the change description and the complete original code, regenerating the entire design without using traceability information.
    \item \textit{Trace-Guided Generation (ours)}: the model takes the change description and trace-linked code fragments to regenerate only the affected snippets and merge them into the original code to obtain the complete updated implementation.
\end{enumerate}

\subsection{Research Questions}

To evaluate the effectiveness of IncreRTL, we design the following research questions (RQs):

\begin{itemize}
    \item \textbf{RQ1:} Can the proposed framework effectively address RTL requirement changes, achieving efficient, consistent, and interpretable regeneration?
    \item \textbf{RQ2:} How does the framework perform when handling diverse types of requirement changes?
    \item \textbf{RQ3:} Can the proposed method maintain consistent effectiveness across different large language models?
\end{itemize}

RQ1 and RQ2 concern the performance and scalability of IncreRTL in handling diverse requirement changes, while RQ3 examines the framework’s generality across different LLMs.

\begin{table}[t]
  \caption{Scalability of IncreRTL across different requirement change types on the EvoRTL benchmark.}
  \label{tab:scale_types}
  \centering
  \footnotesize
  \resizebox{0.5\textwidth}{!}{%
    \begin{tabular}{llccc}
      \toprule
      \textbf{Magnitude} & \textbf{Category} & \textbf{Consistency} & \textbf{Syntax} & \textbf{Function}  \\
      \midrule
      \multirow{4}{*}{\textit{Minor}} 
        & FunctionalBehaviorChange & 0.7935 & 94.12 & 68.24 \\
        & ControlConfigChange   & 0.6846 & 89.66 & 62.07 \\
        & InterfaceProtocolChange  & 0.8109 & 71.43 & 42.85 \\
        & MicroArchitectureRefactor        & 0.7192 & 89.00 & 60.00 \\
      \addlinespace[2pt]
      \midrule
      \multirow{2}{*}{\textit{Major}} 
        & ModuleStructuralRefactor & 0.6142 & 78.26 & 44.35 \\
        & SystemInterfaceRefactor  & 0.7070 & 91.43 & 42.86 \\
      \bottomrule
    \end{tabular}%
  }
\end{table}

\subsection{Experimental Results and Analysis}

\subsubsection{Results for RQ1}
To address RQ1, we compare IncreRTL with two baseline methods on EvoRTL-Bench. As shown in Table~\ref{tab:RQ1}, IncreRTL attains the highest consistency score, outperforming both the Full and Direct baselines. Consistency is particularly critical in hardware design, where requirements frequently evolve, as it ensures that implementation behavior remains stable and predictable across multiple updates. 
In terms of token usage (RTU), Direct Generation consumes the fewest tokens. However, compared with the Full baseline, IncreRTL reduces relative token usage from 1.8× to 1.46× that of Direct, while achieving substantially higher consistency, offering a better trade-off between cost and quality. Although Full exhibits a modest advantage in correctness (a 0.10\% gain in syntax accuracy and approximately a 6\% improvement in functional correctness as measured by pass@5 and pass@10), this benefit is offset by substantially higher token usage and longer inference latency. The broader context window used by Full improves global understanding of the design but introduces considerable computational overhead, which is undesirable for large-scale or iterative design workflows.

By contrast, the Direct baseline performs the weakest because it lacks contextual information, leading to unstable behavior and lower accuracy. IncreRTL’s efficiency and consistency stem from traceability-guided generation, which confines edits to implementation regions directly affected by requirement changes. This localized design limits unnecessary modifications, preserves syntactic structure via hierarchical chunking, and maintains correctness without heavy reliance on long-range context. As designs grow more complex, these advantages become more pronounced, enabling IncreRTL to achieve a more favorable balance among consistency, efficiency, and correctness than both baselines.


\begin{figure}[t]
  \centering
  \includegraphics[width=1\linewidth]{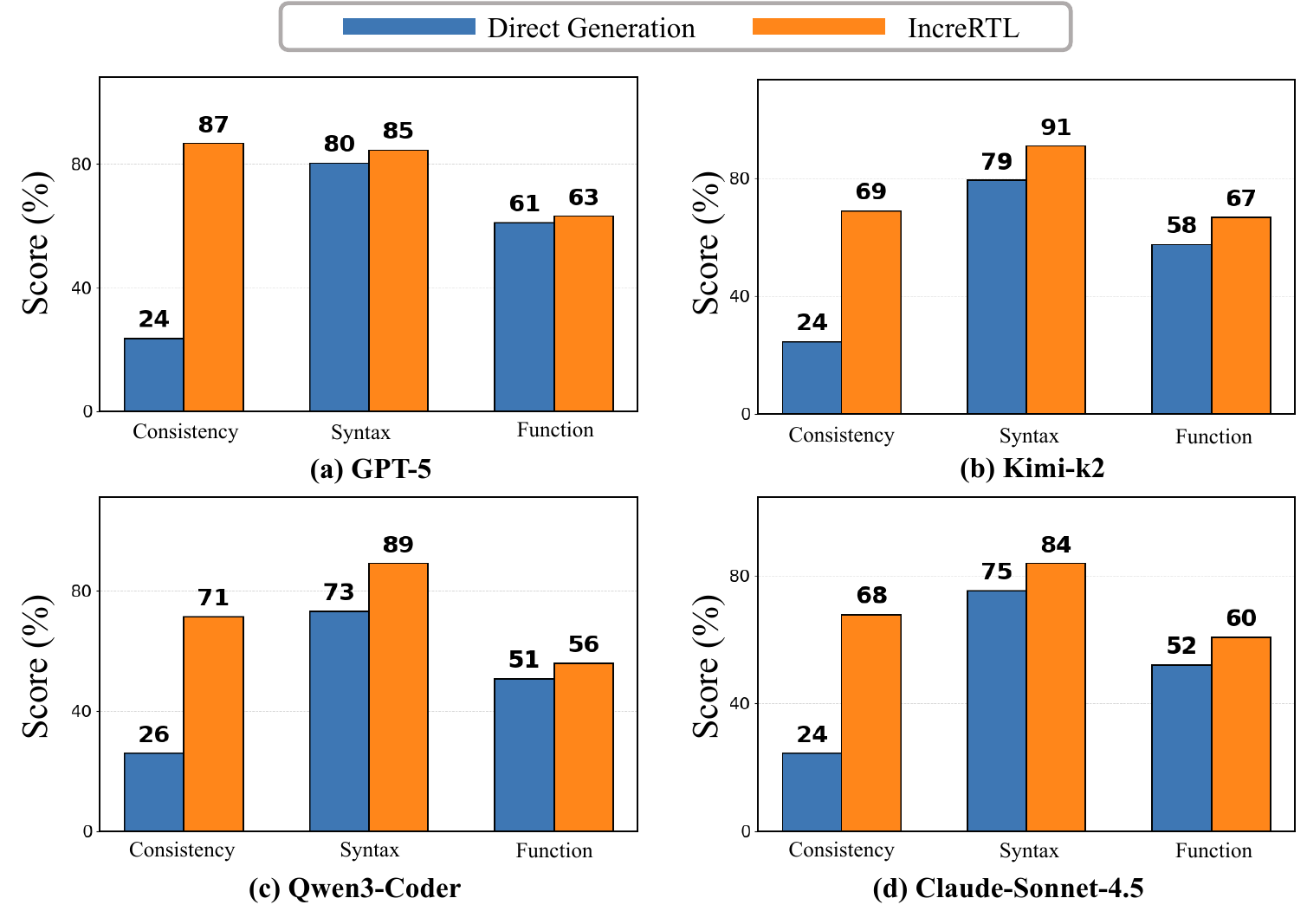}
  \caption{Performance comparison between IncreRTL and Direct Generation across different large models.}
  \Description{Illustration showing requirement-driven modification in LLM-based RTL generation, including standard and evolved workflows.}
  \label{fig:example_merge}
\end{figure}

\subsubsection{Results for RQ2}

To address RQ2, we evaluate the scalability of \textbf{IncreRTL} on the \textbf{EvoRTL benchmark}, which covers six categories of requirement changes (four minor and two major), using the same evaluation metrics as in RQ1. As shown in Table~\ref{tab:scale_types}, IncreRTL maintains strong and stable performance across most minor change types. Specifically, \textit{Functional Behavior Change} achieves high consistency (0.7935), syntax accuracy (94.12\%), and a functional pass rate of 68.24\%. \textit{Control Configuration Change} likewise exhibits stable generation quality, as IncreRTL’s localized update mechanism naturally targets the functional and control regions affected by such modifications. \textit{MicroArchitecture Refactor} remains robust under local optimization changes, whereas \textit{Interface Protocol Change} shows a drop in syntax accuracy (71.43\%) and functional correctness (42.85\%) despite reasonable consistency (0.8109), reflecting the increased difficulty of preserving global consistency under complex inter-module dependencies.

For major changes, IncreRTL experiences a moderate performance decline. Both \textit{Module Structural Refactor} and \textit{System Interface Refactor} maintain relatively good consistency, with scores of 0.6142 and 0.7070, and functional pass rates around 44\%. This reflects the increased difficulty of maintaining generation stability under large-scale dependency propagation. Nevertheless, IncreRTL achieves syntax accuracy of 78.26\% and 91.43\%, indicating its ability to preserve syntactic integrity when handling substantial structural modifications.
 Overall, these results demonstrate the scalability of IncreRTL across requirement changes of different magnitudes.

\subsubsection{Results for RQ3}


To address RQ3, we conducted an evaluation of the generalizability of IncreRTL across various large language models, including GPT-5, Kimi-k2~\cite{team2025kimi}, Qwen3‑Coder and Claude-Sonnet-4.5, using the same evaluation metrics as in prior experiments. The results show that stronger models achieve higher accuracy for the Direct Generation baseline, particularly in terms of syntax and functional correctness, as their larger capacity enables more precise code. However, even with these stronger models, the Direct baseline still exhibits low consistency scores, which implies higher testing costs and debugging effort. As illustrated in Figure~\ref{fig:example_merge}, adding IncreRTL on top of each model substantially improves consistency while also enhancing correctness. By leveraging requirement–to–code traceability links, IncreRTL stabilizes the generation process across different models, demonstrating strong generalizability.



\section{Conclusion}
This study provides the first systematic analysis of the challenges in adapting LLM-based RTL code generation to evolving design requirements. We introduce a requirement traceability-driven framework, accompanied by a benchmark and evaluation metrics. Experimental results highlight the proposed approach's superior consistency and efficiency over existing methods. As RTL design complexity grows, requirement-aware adaptation is poised to become a critical research focus.

\clearpage

\bibliographystyle{ACM-Reference-Format}
\bibliography{sample-base}


\end{document}